\documentclass{aa}  
\usepackage{aalongtable,lscape}
\usepackage{graphicx}
\usepackage{txfonts}
\begin{document}
\title{Stars and brown dwarfs in the $\sigma$ Orionis cluster: \\
the Mayrit catalogue} 
\titlerunning{Stars and brown dwarfs in $\sigma$ Orionis}
%
%
\author{Jos\'e A. Caballero\inst{1}\fnmsep\thanks{{\em Formerly:} Alexander
von Humboldt Fellow at the Max-Planck-Institut f\"ur Astronomie.
{\em Currently:} Investigador Juan de la Cierva at the Universidad Complutense
de~Madrid.}}
%
%
\institute{Max-Planck-Institut f\"ur Astronomie, K\"onigstuhl 17, D-69117
Heidelberg, Germany, \email{caballero@astrax.fis.ucm.es}}
\date{Received May .., 2007; accepted October 25, 2007}

\abstract
{The young $\sigma$~Orionis cluster is an indispensable basis for understanding
the formation and evolution of stars, brown dwarfs and planetary-mass objects. 
Our knowledge of its stellar population is, however, incomplete.}
{I present the Mayrit catalogue, that comprises most of the stars and high-mass
brown dwarfs of the cluster.}  
{The basis of this work is an optical-near infrared correlation between the
2MASS and DENIS catalogues in a circular area of radius 30\,arcmin centred on
the OB-type binary $\sigma$~Ori~AB.
The analysis is supported on a bibliographic search of confirmed cluster members
with features of youth and on additional X-ray, mid-infrared and astrometric
data.}    
{I list 241 $\sigma$~Orionis stars and brown dwarfs with known features of
youth, {97} candidate cluster members (40 are new) and {115} back- and
foreground sources in the survey area.
The {338} cluster members and member candidates constitute the Mayrit
catalogue.}  
{This catalogue is a suitable input for studying the spatial ditribution,
multiplicity, properties and frequency of discs and the complete mass function
of $\sigma$~Orionis.}  
\maketitle
%

\section{Introduction}

The $\sigma$ Orionis cluster in the \object{Ori~OB~1b} Association is getting as
important for the study of the formation, evolution and characterisation of
stars and substellar objects as other famous clusters and star-forming regions,
like the \object{Hyades}, the \object{Pleiades}, the \object{Orion Nebula
Cluster} or the \object{Taurus-Auriga Complex}.  
The $\sigma$~Orionis cluster is young (3$\pm$2\,Ma), nearby (
$\sim$385\,pc) and relatively free of extinction (Lee 1968; Brown et~al.
1994; Oliveira et~al. 2002; Zapatero Osorio et~al. 2002a; Sherry 
et~al. 2004; B\'ejar et~al. 2004b; Caballero 2007d). 
Firstly identified by Garrison (1967) and Lyng\aa~(1981), $\sigma$~Orionis was
rediscovered by Wolk (1996) and Walter et~al. (1997).
They reported a clustering of young low-mass stars, many of them positionally
coincident with X-ray sources, surrounding the Trapezium-like, multiple stellar
system $\sigma$~Ori in the vicinity of the \object{Horsehead Nebula} (see
Caballero 2007b for a description of the multiple system $\sigma$~Ori and its
surroundings).
Previously, the area had been investigated during wide searches in the Orion
complex with prism-objective and Schmidt plates, detecting a wealth of emission
stars (e.g. Haro \& Moreno 1953; Wiramihardja et~al. 1989), but the cluster had
not been treated as an independent entity within the complex. 
Complete compilations of the determinations in the literature of the age,
heliocentric distance, frequency of discs and mass function of the
$\sigma$~Orionis cluster are in Caballero (2007a). 
A chapter of the Handbook of Star Forming Regions, edited by B.~Reipurth, will
be exclusively devoted to $\sigma$~Orionis (F.~M.~Walter et~al., 
in~press).

After the seminal work by B\'ejar et~al. (1999), who found
for the first time a rich population of young brown dwarfs in $\sigma$~Orionis,
the cluster has turned into an excellent laboratory for the study of, e.g.: 

\begin{itemize}
\item the search for free-floating planetary-mass objects (with masses below the
deuterium-burning limit) and the study of the substellar mass function down to a
few Jupiter masses (Zapatero Osorio et~al. 2000; B\'ejar et~al. 2001;
Gonz\'alez-Garc\'{\i}a et~al. 2006; Caballero et~al. 2007);
\item the frequency and the properties of $\sim$3\,Ma-old discs at different
mass intervals (Jayawardhana et~al. 2003; Oliveira et~al. 2004, 2006;
Hern\'andez et~al. 2007; Caballero et~al. 2007; Zapatero Osorio et~al.
2007a);
\item the masses of OB-type stars in resolved binary systems (Heintz 1997; Mason
et~al. 1998; Caballero 2007d);
\item the X-ray emission of young stars and brown dwarfs (Mokler \& Stelzer
2002; Sanz-Forcada et~al. 2004; Franciosini et~al. 2006; Caballero 2007b);
\item the characterisitics of jets and Herbig-Haro objects (Reipurth et~al.
1998; L\'opez-Mart\'{\i}n et~al. 2001; Andrews et~al. 2004); and
\item the photometric variability of low-mass stars and brown dwarfs
(Bailer-Jones \& Mundt 2001; Caballero et~al. 2004; Scholz \& Eisl\"offel~2004).
\end{itemize}

Many interesting star-like objects have been discovered in the cluster, from the
helium-rich, B2.0Vp-type magnetic star $\sigma$~Ori~E (Greenstein \& Wallerstein
1958), through the Class~I object candidate IRAS~05358--0238 (Oliveira \& van
Loon 2004), to the hypothetical proplyd $\sigma$~Ori~IRS1 (van Loon \& Oliveira
2003; Caballero 2005, 2007b). 
The most interesting objects in the cluster are, however, below the
hydrogen-burning mass limit.
Some of these substellar objects are 
the $\sim$T6-type object \object{S\,Ori~70} (which may be the least
massive body directly detected out of the Solar System, $\sim$3\,$M_{\rm Jup}$
-- Zapatero Osorio et~al. 2002c, 2007b; Burgasser et~al. 2004), the T
Tauri substellar analog S\,Ori~J053825.4--024241 (which is the most variable
brown dwarf yet found; Caballero et~al. 2006a), the two strong H$\alpha$
emitters at the planetary boundary \object{S\,Ori~55} and \object{S\,Ori~71}
(with masses of only 10--20\,$M_{\rm Jup}$ and equivalent widths of the
H$\alpha$ line of up to --700\,\AA; Zapatero Osorio et~al. 2002b; Barrado y
Navascu\'es et~al. 2002a) and the brown dwarf-exoplanet system candidate
\object{SE~70} + \object{S\,Ori~68} (which could be the widest planetary system
known so far; Caballero et~al. 2006b). 
The number of known substellar objects in $\sigma$~Orionis is comparable to
those of other rich, more massive, younger star-forming regions like Chamaeleon,
Ophiuchus or the Orion Nebula Cluster.
However, $\sigma$~Orionis is by far the region with the largest amount of brown
dwarfs with membership confirmation ($>$ 30; Caballero et~al. 2007) and
planetary-mass object candidates (29; Zapatero Osorio et~al. 2000;
Gonz\'alez-Garc\'{\i}a et~al. 2006; Caballero 2007b; Caballero et~al. 2007).
Many of the latter bodies have measured L spectral types and/or flux excess
longwards of 5\,$\mu$m (Zapatero Osorio et~al.~(2007a).

Important efforts have been recently carried out to characterise the
$\sigma$~Orionis cluster in general and to investigate the connection
between its stellar and substellar populations in particular (e.g. B\'ejar
et~al. 2004a; Sherry et~al. 2004; Kenyon et~al. 2005; Burningham et~al. 2005;
Caballero 2005, 2007a, 2007b, 2007c). 
The works by Kenyon et~al. (2005), who investigated membership, binarity and
accretion among very low-mass stars and brown dwarfs surrounding $\sigma$~Ori,
and, especially, Sherry et~al. (2004) stand out.
The latter authors presented an ambitious study, estimating the number of
cluster members in the mass range 0.2\,$M_\odot$ $\lesssim M
\lesssim$ 1.0\,$M_\odot$ and the radius, age and total mass of the cluster.
All these works are, however, incomplete or biased in some way: no
comprehensive, homogenous, multi-band study, fully covering the whole stellar
mass interval (from $\sim$20\,$M_\odot$ down to the substellar boundary) and the
cluster area without gaps (from the very centre to the border) exists so far in
$\sigma$~Orionis.

I extend the study of the brightest stars of the cluster shown
in Caballero (2007a) down to the hydrogen-burning mass limit and beyond.
I start out from a correlation between the 2MASS and DENIS catalogues 
within the environment of the Virual Observatory\footnote{See
{http://www.ivoa.net}.} and a bibliographic search of confirmed cluster
members, and support them with spectroscopic and photometric data from the X-ray
region to 120\,$\mu$m when available.
The outcome of this work is the Mayrit catalogue, that comprises the
majority of the stars and high-mass brown dwarfs of $\sigma$~Orionis.

\section{Analysis and results}

\subsection{The 2MASS/DENIS correlation}
\label{correlation}

\begin{figure}
\centering
\includegraphics[width=0.49\textwidth]{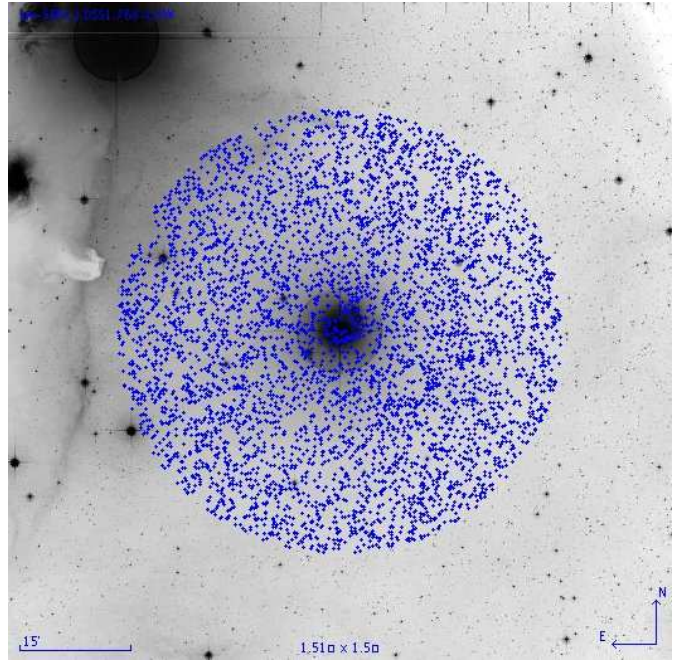}
\caption{Inverse-colour DSS-1 blue-band (photographip $B_J$) image centred in
$\sigma$~Ori~AB.
Its size is 1.5 $\times$ 1.5\,deg$^2$.
North is up and east is left.
The correlated 2MASS/DENIS sources are marked with (blue) dots.
Compare this Figure with fig.~1 in Caballero (2007a).
Colour versions of all the Figures are available in the electronic
publication.} 
\label{aladin}
\end{figure}

The correlation between the Two-Micron All Sky Survey (2MASS) Catalog of
Point Sources (Skrutskie et~al. 2006) and the third release of the 
Deep Near Infrared Survey of the Southern Sky (DENIS) database (Epchtein
et~al. 1997; DENIS Consortium 2005) were performed with the cross match tool of
the Aladin sky atlas.
I~used a cross-match threshold of 1\,arcsec, which is more than 10 times wider
than the expected astrometric errors of the catalogues.
The methodology and the survey area, a circle of 30\,arcmin radius centred on
$\sigma$~Ori~AB, are identical to those in Caballero (2007a), where further
details can be found.
The survey area and its surroundings are shown in Fig.~\ref{aladin}.

Of the {5721} 2MASS sources in the 2830\,arcmin$^2$-wide area, 4951
($\sim$87\,\%) were correlated with DENIS sources.
Most of the 770 non-correlated 2MASS sources are very faint and are
expected to have optical counterparts fainter than the DENIS completeness.
To avoid subsequent problems associated to high photometric uncertainties (e.g.
faint blue sources with apparent red colours and vice versa), I~only considered
2MASS sources with photometric errors $\delta K_{\rm s} <$ 0.1\,mag.
With this restriction, the final sample size was 2332 optical/near-infrared
sources.  

\begin{figure}
\centering
\includegraphics[width=0.50\textwidth]{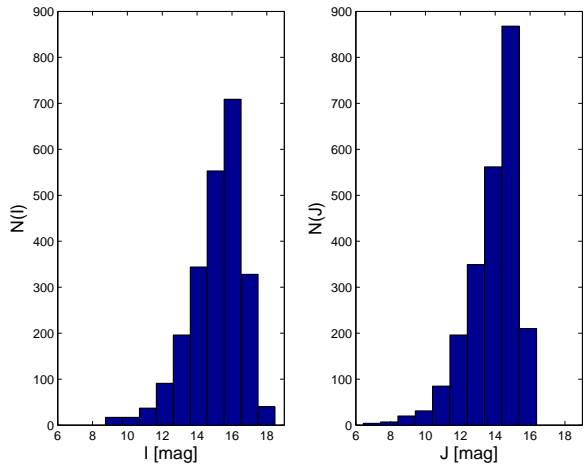}
\caption{Histograms of the number of sources as a function of DENIS $I$ ({\em
left}) and 2MASS $J$ magnitudes ({\em right}).}
\label{NIINJJ}
\end{figure}

The histograms of the number of sources as a function of magnitude in the DENIS
$I$ and 2MASS $J$ passbands are shown in Fig.~\ref{NIINJJ}.
The bulk of the correlated sources have $I$- and $J$-band magnitudes in the
intervals $\sim$12--17\,mag and $\sim$11--16\,mag, respectively.
The number of sources peaks at $I \sim$ 16\,mag and $J \sim$ 15\,mag (and at $H
\sim K_{\rm s} \sim$ 14\,mag), which is expected from the increasing number of
faint stars and the completeness and limiting magnitudes of the DENIS and 2MASS
catalogues (DENIS $I_{5\sigma}$ = 18.0\,mag; 2MASS $J_{3\sigma}$ = 17.1\,mag,
$K_{{\rm s},3\sigma}$ = 14.3\,mag). 
Both surveys are complete at magnitudes brighter than $I \approx$ 16.5\,mag, $J
\approx$ 14.6\,mag and $K_{\rm s} \approx$ 12.8\,mag, which are the
magnitudes of the (non-reddened) faintest stars and brightest brown dwarfs
in $\sigma$~Orionis (Caballero et~al. 2007).
Besides, 98\,\% (91\,\%) of the correlated sources have $I-J$ colours in
the interval 0.4\,mag $\lesssim I-J \lesssim$ 2.5\,mag (0.7\,mag $\lesssim I-J
\lesssim$ 2.1\,mag), which are typical of normal dwarfs.
Since the galactic latitude of the cluster is $b$ = --17.3\,deg, a large
contamination by red giants is not expected (Caballero, Burgasser \&
Klement, in~prep.). 

\begin{figure}
\centering
\includegraphics[width=0.50\textwidth]{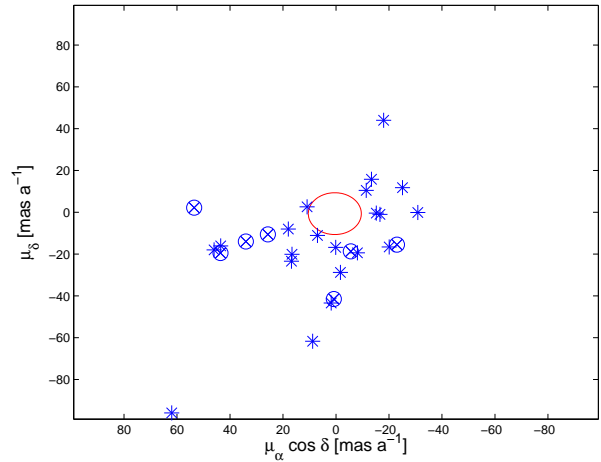}
\caption{Proper-motion diagram ($\mu_\delta$ vs. $\mu_\alpha \cos{\delta}$) of
foreground stars with $\mu >$ 10\,mas\,a$^{-1}$.
Non-cluster members from the literature and new non-cluster members from this
work are shown with asterisks (``$\ast$'') and tensorial product symbols
(``$\otimes$''), respectively. 
The threshold used by Caballero (2007a) to separate Ori~OB~1b members from stars
with high tangential velocities is marked with a big ellipse.
The high-proper-motion star G~99--20 (LP 598--162; $\mu$ = 288\,mas\,a$^{-1}$)
is not shown for the sake of clarity.} 
\label{muramude}
\end{figure}
%

Other optical catalogues offered by the Virtual Observatory that could have
been correlated with 2MASS are, for example, the United States Naval Observatory
USNO-B1.0 (Monet et~al. 2003), the SuperCOSMOS Sky Survey SSS (Hambly et~al.
2001), the  Guide Star Catalogue\footnote{Space Telescope
Science Institute and Osservatorio Astronomico di Torino.} GSC-II or the
Carlsberg Meridian Catalog\footnote{Copenhagen University Obs., Institute of
Astronomy, Cambridge, UK, and Real Instituto y Observatorio de la Armada en San
Fernando} CSC14. 
All of them are digitisations of Schmidt photographic plates.
In spite of possible photometric uncertainties (see Section~\ref{refinement}),
the DENIS catalogue is deeper than them.
Besides, it is a CCD-based survey, which easily allows to transform magnitudes
into fluxes and warranties linearity within the dynamical range of the survey.
Photographic magnitudes, on the contrary, are not reliable at the faintest
(especially GSC-II and CSC14) and brightest (especially SSS\footnote{The
SSS/2MASS correlation gives rather blue optical/near-infrared colours, by up to
2\,mag ($I_{\rm SSS} - K_{\rm s}$), with respect to observational (Bessel \&
Brett 1988) and theoretical (Schaller et~al. 1992; Baraffe et~al. 1998; Girardi
et~al. 2000) star sequences for all objects brighter than $I \approx$
14\,mag.}) magnitudes.
An optimum alternative to DENIS would be using USNO-B1 and SSS at the
faintest and brightest magnitudes, respectively.
This option still has an important drawback with respect to DENIS (apart of a
complex implementation): the incomplete coverage of the cluster centre due to
the glare of $\sigma$~Ori~AB. 
The inner 3\,arcmin-radius circle contains about 40 cluster members and
candidates and an important fraction of the cluster total mass (Caballero 2007a,
2007b).
On the one hand, DENIS tabulates optical data for {\em all} the sources with
near-infrared magnitudes $J \lesssim$ 16.0\,mag from the outer radius of the
survey at 30\,arcmin to a few arcseconds to the Trapezium-like star system;
more than 60 DENIS sources are in the central circle of radius 3\,arcmin.
On the other hand, USNO-B1 tabulates only {\em three} sources in the central
circle (the OB-type stars $\sigma$~Ori~AB, D and E), and is complete and
relatively free of glare artifacts only at more than $\sim$6\,arcmin. 
The effect of ``glare incompleteness'' is also detectable surrounding other
bright early-type stars in the area with known nearby companions (e.g.
HD~294271; Caballero~2005). 

There are other non-virtual observatory optical data covering the 
$\sigma$~Orionis region, but they are not available to the author (e.g. Sherry
et~al. 2004), are too deep (i.e. all high- and intermediate-mass stars are
saturating; e.g. B\'ejar et~al. 2004a, 2004b) or on purpose avoided to survey
the cluster centre (e.g. Caballero 2006).
To sum up, the DENIS/MASS combination provides the most suitable
optical/near-infrared correlation in the~area; 
this combination is, nevertheless, incomplete for magnitudes brighter than $I
\sim$ 6\,mag and offers incorrect magnitudes of objects brighter than $I
\approx$ 10\,mag due to saturation and non-linear effects.
To enhance the optical data from the DENIS catalogue, I~have collected the $I$
magnitudes of the 18 brightest stars in the area from the USNO-B1 catalogue and
from Caballero (2007b).  
Likewise, the 2MASS $J$ and $H$ photometry of $\sigma$~Ori~AB is also affected
by saturation, and I~have used the respective values provided by Johnson et~al.
(1966).
These incorporations leaded to simultaneously investigate in the survey all the
sources with $J$-band magnitudes in the range 4.5--15.5\,mag.
Therefore, all the cluster stars and many of the high-mass brown dwarfs can be
identified in the optical/near-infrared~data.

\subsection{Membership classification}

\subsubsection{Known cluster members}
\label{knownclustermembers}

As a first step of the cluster member selection, a list of confirmed cluster
members from the literature with 2MASS/DENIS counterpart was made up, given in
Table~\ref{spektra}. 
A confirmed cluster member must display at least one of the following features
of youth, characteristic of $\sim$3\,Ma-old stars or brown dwarfs: 
\begin{itemize}
\item early spectral type (``OB''): O-, B- and early-A-type stars taken from
Caballero (2007a);  
\item Li {\sc i} $\lambda$6707.8\,\AA~in absorption (``Li {\sc i}''): mid- and
late-type pre-main sequence stars mostly taken from Wolk (1996), Zapatero
Osorio et~al. (2002a), Kenyon et~al. (2005) and Caballero (2006);  
\item strong and/or broad H$\alpha$ $\lambda$6562.8\,\AA~line in emission
(``H$\alpha$''): accretors and emission stars and brown dwarfs mostly taken from
Haro \& Moreno (1953; Haro objects), Wiramihardja et~al. (1989, 1991; Kiso
objects), Zapatero Osorio et~al. (2002a), Weaver \& Babcock (2004) and Caballero
(2006); 
\item features of low gravity (``low $g$''): stars and brown dwarfs with
weak alkali absorption lines (in particular, the Na {\sc i}
$\lambda\lambda$8183,\,8195\,\AA~doublet) taken from Burningham et~al. (2005);
\item spectral energy distribution characteristic of objects with discs (``Class
I'', ``II'', ``trans. disc'', ``ev. disc'', ``mIR''): Class I and Class II
objects and stars with evolved or transition discs mostly taken from Oliveira
\& van Loon (2004) and Hern\'andez et~al. (2007);
\item very strong X-ray emission (``XX'', ``XXX''): optical/near-infrared
counterparts of: ($i$) X-ray sources with count rates in the intervals
0.01--0.10\,s$^{-1}$ (``XX'') or $>$0.10\,s$^{-1}$ (``XXX'') in the EPIC/{\em
XMM-Newton} observations by Franciosini et~al. (2006), or of ($ii$) X-ray
sources out of the EPIC/{\em XMM-Newton} survey area detected by at least two
independent surveys carried out with the space observatories {\em Einstein},
{\em ASCA} and {\em ROSAT} and with count rates $>$0.01\,s$^{-1}$ in the
WGACAT/{\em ROSAT} catalogue, mostly taken from Harris et~al. (1994), Nakano et
al. (1999) and White et~al. (2000).
\end{itemize}

Table~\ref{spektra} provides the Mayrit identifications of the 241
confirmed young stars and brown dwarfs in $\sigma$~Orionis that I was able to
identify among the correlated 2MASS/DENIS sources.
It also gives some of their alternative names from the literature, their
features of youth and corresponding abbreviated bibliographic references.  
The complete reference list is given in Table~\ref{references}.
Some additional features of youth are also indicated in Table~\ref{spektra}
(``Em.'', ``Ca {\sc ii}'': emission lines different from H$\alpha$; ``Si'':
silicon features in mid-infrared spectra, characteristic of dust coagulation in
protoplanetary discs; 
``debris disc'': flux
excess detected only at the 24\,$\mu$m passband of MIPS/{\em Spitzer};
``X-flare'': flares detected at 0.2--2.0\,keV; ``jet'': central source
associated to a Herbig-Haro object; ``nebular'': nebular appearence of the
H$\alpha$ emission; ``X'': stars with EPIC/{\em XMM-Newton} count rates in the
interval 0.0005--0.01\,s$^{-1}$). 
Some marginally-detected features are marked with a question mark.
Of the 241 identified cluster members, 187 have at least two known
features of youth (and 95 have three or more features). 
Stars and brown dwarf candidates with only spectral-type determination from
low-resolution spectroscopy have not been considered as confirmed
cluster~members.

There are also several cluster member candidates, like
\object{Kiso~A--904~69} ([WB2004]~17), \object{Haro~5--28} (Kiso~A--0904~89),
\object{IRAS~05352--0227} or \object{Haro~5--24}, with published strong
H$\alpha$ emission and magnitudes that would make them to be detectable in
this survey.
However, because of the large errors in their coordinates, I was not able
to identify~them.  

The Mayrit designation of the $\sigma$~Orionis cluster members and
candidates used throughout this work follows the nomenclature introduced by
Caballero~(2007b).

\subsubsection{Fore- and background stars and galaxies}
\label{nonmembers}

\begin{figure*}
\centering
\includegraphics[width=0.80\textwidth]{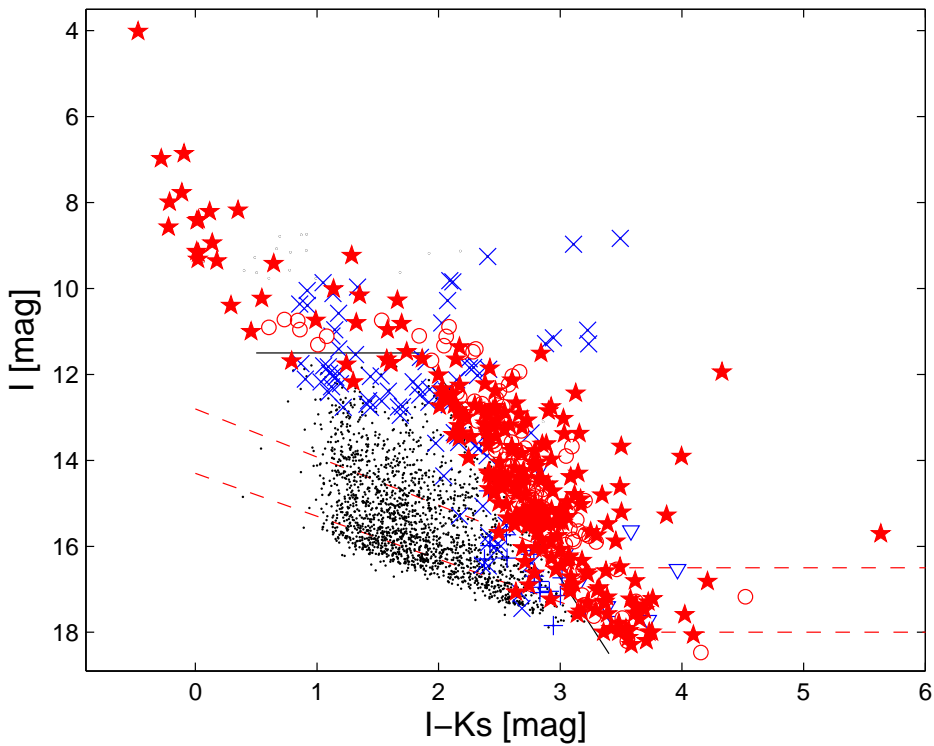}
\caption{$I$ vs. $I-K_{\rm s}$ colour-magnitude diagram in the survey area.
Filled (red) stars: stars and brown dwarfs with known features of youth from the
literature;
open (red) circles: photometric candidate cluster stars and brown dwarfs;
(blue) pluses: galaxies;
open (blue) down triangles: reddened sources in the northeastern cloud;
(blue) x-marks: fore- and background stars;
tiny (black) dots: unresolved galaxies, probable fore- and background stars and
cluster stars and brown dwarfs with blue $I-K_{\rm s}$ colours for their $I$
magnitudes;
solid (black) line: criterion for selecting candidate cluster stars and brown
dwarfs without known features of~youth;
dashed (red) lines: approximate completeness and detection limits of the
survey.} 
\label{IKsI}
\end{figure*}
%

\begin{figure*}
\centering
\includegraphics[width=0.80\textwidth]{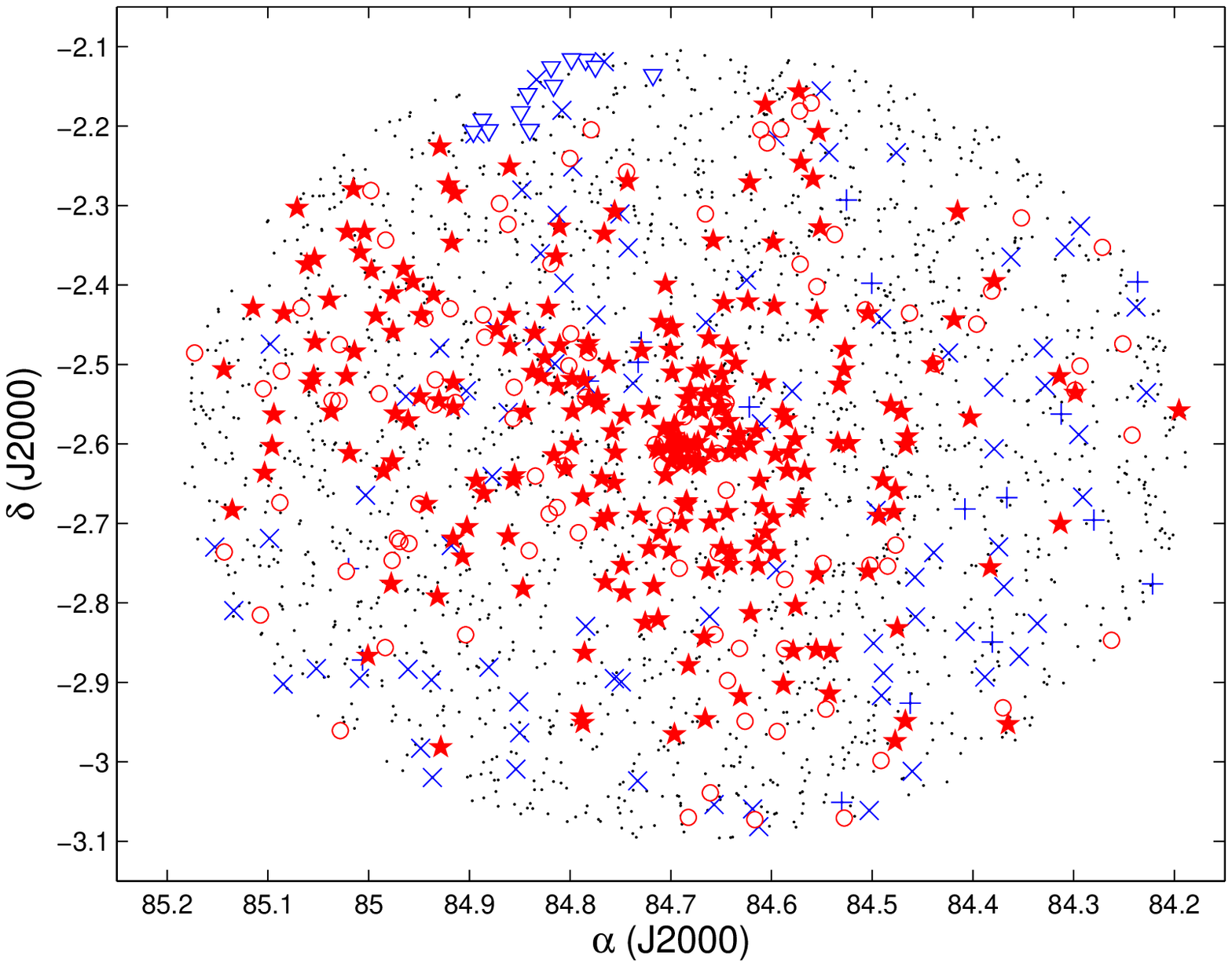}
\caption{Spatial distribution of the investigated sources.
The symbol code is as in Fig.~\ref{IKsI}.}
\label{RADE}
\end{figure*}
%

To complement the list of known cluster members, I compiled 85
non-cluster member stars and 17 galaxies among the correlated
2MASS/DENIS sources, shown in Table~\ref{NM}. 
Nine stars are presented here for the first time, while the remaining
non-cluster members were taken from the literature.
In particular, the latter stars are:
\begin{itemize}
\item bright stars with high tangential velocity and K- and M-type foreground
stars taken from Caballero (2007a); and
\item stars whose high-quality spectra show no lithium in absorption (``no Li
{\sc i}''), no low gravity features in the alkali lines (``no low $g$'') and/or
discordant radial velocity with respect to the cluster systemic radial velocity
(``no V$_r$''), taken from Kenyon et~al. (2005), Burningham et~al. (2005) and
Caballero (2006). 
\end{itemize}
I~also got the proper motion tabulated by USNO-B1 (very similar to those
tabulated by SSS and NOMAD1 -- Zacharias et~al. 2004) for all the
correlated 2MASS/DENIS sources brighter than $I$ = 12.5\,mag that were not
investigated in the Tycho-2/2MASS correlation by Caballero (2007a). 
I~estimate that the average error of the USNO-B1 measurements in the
investigated magnitude interval is $\delta \mu \gtrsim$ 5\,mas\,a$^{-1}$.
There are, however, some stars with relatively high proper motions and
larger uncertainties (e.g. SO430116, with the second highest proper motion in
the area after G~99--20, has a SSS $\mathbf \mu$ =
(+103$\pm$5,--123$\pm$5)\,mas\,a$^{-1}$, while USNO-B1 tabulates a proper motion
$\sim$30\,\% less). 
Of the 85 non-cluster members, 29 have proper-motion moduli larger than
10\,mas\,a$^{-1}$, which was the value to separate young stars and candidate
Ori~OB~1b association members from probable foreground stars with larger
tangential velocities in Caballero (2007a).
The proper-motion diagram shown in Fig.~\ref{muramude} illustrates the study of
the star tangential velocities.
Seven of the nine new non-cluster members firstly listed here have proper
motions $\mu \gtrsim$ 20\,mas\,a$^{-1}$.
The remaining two new non-cluster members, with proper motions consistent with
membership in $\sigma$~Orionis, probably are foreground K- or early-M-type stars
because they have very red colours, but they are much brighter than cluster
members of similar expected spectral type and do not show any feature
characteristic of T~Tauri stars.
These two stars are \object{2MASS~J05400217--0253423} ($I$ =
10.97$\pm$0.06\,mag, $K_{\rm s}$ = 7.74$\pm$0.03\,mag) and
\object{2MASS~J05390143--0253431} ($I$ = 11.28$\pm$0.03\,mag, $K_{\rm s}$ =
8.40$\pm$0.03\,mag).
Table~\ref{NM} gives the 2MASS designations of all the fore- and background
stars, alternative names if they exist, a flag indicating if they have been
spectroscopically studied (``Sp.'' = Yes/No), their USNO-B1 proper motions
if they are larger than 10\,mas\,a$^{-1}$, remarks and abbreviated references
(see again Table~\ref{references} for complete references). 
The nine new non-cluster members have a blank in the references column.

In Table~\ref{galaxies}, I list the 2MASS designations of 18
galaxies found in the survey area. 
All except one (2E~1448, which is a very strong X-ray emitter; Caballero \&
L\'opez-Santiago, in~prep.) are tabulated in the Final Release of 2MASS
Extended Sources (2MASS 2003).
Some of galaxies were also identified in the {\em Spitzer} Space Telescope study
in $\sigma$~Orionis by Hern\'andez et~al. (2007).
As an additional test, I~checked the extended FWHMs of the ten
reddest galaxies ($J-K_{\rm s} \gtrsim$ 1.4\,mag) using the 
SSS~digitisations.

\subsubsection{Candidate young stars and brown dwarfs}
\label{selection}

I~have selected {109} cluster member candidates without membership
information based on their position with respect to the 241 confirmed
cluster members in the $I$ vs. $I-K_{\rm s}$ colour-magnitude diagram shown in
Fig.~\ref{IKsI}. 
The cluster members with known features of youth define a quite broad
photometric sequence in the diagram. 
I~have classified as candidate cluster stars and brown dwarfs those correlated
2MASS/DENIS sources that are neither known cluster members
(Section~\ref{knownclustermembers}) nor non-cluster members
(Section~\ref{nonmembers}), and that fall redwards of the solid line shown in
Fig.~\ref{IKsI} or are brighter than $I$ = 11.5\,mag. 
The solid line leaves redwards of it $\sim$75\,\% of the confirmed cluster
members.
All the stars in the area brighter than $I$ = 9.5\,mag are confirmed
cluster members. 

\begin{figure}
\centering
\includegraphics[width=0.50\textwidth]{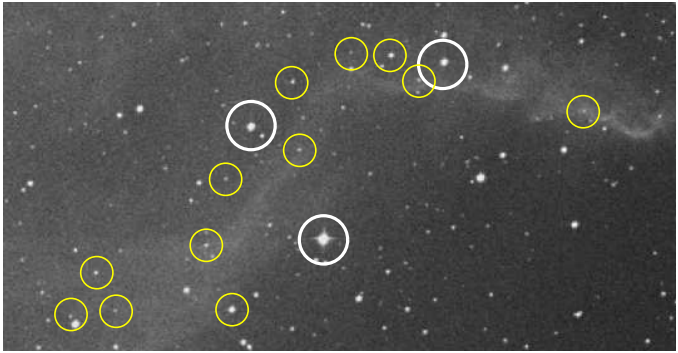}
\caption{Part of the ESO Red (photographic $R$) image of the northeastern
nebulosity discussed in the text.
The three known non-cluster members are marked with big thick (white) circles.
The 12 reddened sources previously considered as candidate
cluster members are marked with small thin (yellow) circles.
The gas emission is due to H$\alpha$ (mostly) and [N {\sc ii}] $\lambda
\lambda$6548.0,6583.4\,\AA.
North is up and east is to the left; approximate size is
17$\times$7\,arcmin$^2$.}
\label{cloud}
\end{figure}

Fig.~\ref{aladin} shows that the survey area partially overlaps with a
nebulosity that forms an extension of the \object{Orion~B} cloud associated to
the \object{Alnitak}-Horsehead Nebula-\object{IC~434}-\object{Flame Nebula}
complex (Alnitak = $\zeta$~Ori). 
Twelve candidate cluster members to the northeast of the survey
area, together with three known non-cluster members with spectroscopic
information, fall exactly below the densest (filament-shaped) part of the
nebulosity. 
This area is shown in detail in Fig.~\ref{cloud} (see also the upper part of
fig.~2 in Hern\'andez et~al. 2007).
The cloud shred also coincides with an alignment of mid-infrared sources in the
{\em IRAS} Catalogue of Point Sources and in the {\em IRAS} Serendipitous Survey
Catalog without known optical or near-infrared counterparts (IPAC 1986; also in
the {\em IRAS} Point Source Reject Catalog). 
Although some of these sources might be bona-fide, embedded young sources
associated to the filament, I list them in Table~\ref{northernrag} as
probable reddened sources that do not belong to the $\sigma$~Orionis cluster. 
Three of them also were photometric cluster member candidates in the work by
Sherry et~al. (2004).
Besides, B\'ejar et~al. (2004a) proposed that an overdensity of $IZ$-band
photometric cluster members to the northeast of $\sigma$~Orionis could be
associated to a hypothetical cluster surrounding Alnitak.
The existence of the nebulosity adds further complexity to the topic.
The solution of this dilemma requires further spectroscopy.

Taking into account the {109} sources redwards of the solid line in
Fig.~\ref{IKsI} and the 12 probable reddened sources in
Table~\ref{northernrag}, then 97 reliable candidate cluster members
without membership information, of which 40 are new, remain from the
survey. 
Their designations and alternative names, together with remarks and
references, are provided in Table~\ref{new}. 
In the remarks column it is indicated, if known, spectral type, period of
photometric variability and X-ray emission (``X'': count rate in the interval
0.0005--0.01\,s$^{-1}$ in Franciosini et~al. 2006; ``X?'': source possibly
associated to X-ray events in the WGACAT/{\em ROSAT} catalogue by White et~al.
2000).  
The marginal detections of youth features (Li {\sc i}, H$\alpha$, low $g$) are
indicated with a question mark.

The pictogramme of the spatial distribution of confirmed cluster members,
non-cluster members, galaxies and candidate cluster members is displayed in
Fig.~\ref{RADE}.
Since most of spectroscopic, mid-infrared and X-ray follow-ups are focused
on the cluster centre, photometric candidate cluster members do not
concentrate towards the cluster centre as sharply as confirmed cluster
members~do\footnote{Caballero (2007c) used a previous version of the Mayrit
catalogue to investigate the spatial distribution of stars and brown dwarfs in
$\sigma$~Orionis. 
The only difference was that the X-ray galaxy 2E~1456 was considered as a
photometric cluster member candidate.
Even accounting for this difference, the results of that paper (clustering
parameter $\mathcal{Q} \approx$ 0.88, volume density $\rho(r) \propto r^{-2}$,
mass-dependent radial distribution, azimuthal asymmetry) remain~unchanged.}. 

%

\subsection{Refinement of optical photometry}
\label{refinement}

As a final step of the analysis, I improved the DENIS $I$ data of the correlated
sources. 
The DENIS database sometimes tabulates two or more detections separated by only
$\sim$0.1--0.3\,arcsec with slightly different magnitudes. 
A visual inspection shows that the detections are associated to the same
source.
Often, one or some detections have very large photometric uncertainties, of up
to $\delta I$ = 1.00\,mag.
During the the 2MASS/DENIS cross match, only one DENIS source (the nearest one
to the 2MASS counterpart) was correlated, which may have larger photometric
errors than the other sources.
Therefore, I investigated the possible multiple detections in the DENIS database
of the 241 confirmed cluster members and the {97} candidate cluster
members. 
Of them, 55 sources ($\sim$16\,\%) have such multiple (double) detections,
$I_1$~and~$I_2$.

\begin{figure}
\centering
\includegraphics[width=0.50\textwidth]{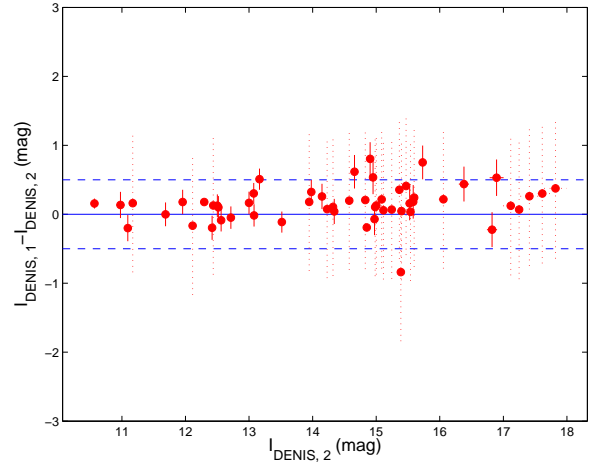}
\caption{$I_1-I_2$ vs. $I_2$ of stars with double detections in the DENIS
database.
The errorbars associated to stars with photometric uncertainties $\delta
I$ = 1.00\,mag are plotted with dotted lines.
Horizontal lines denote $I_1-I_2$ = 0.0\,mag (solid) and $I_1-I_2$ =
$\pm$0.5\,mag (dashed).}
\label{III}
\end{figure}

The difference between the two tabulated I-band magnitudes, $I_1$
and $I_2$, as a function of the most accurate measurement is shown in
Fig.~\ref{III}. 
While the mean of the quantity $(\delta^2 I_1 + \delta^2 I_2)^{1/2}$, that
measures the combined photometric uncertainties, is 0.56\,mag, the standard
deviation of the difference $I_1-I_2$ is only 0.26\,mag.
There are only three objects with $I_1-I_2 >$ 0.56\,mag 
(V603~Ori, Haro5--18 and SWW~34),  
of which two are known to be intense accretors and one is known to be a
photometric variable (Haro \& Moreno 1953; Salukvadze 1987;
Wiramihardja et~al. 1989; Andrews et~al. 2004). 
These values must be compared with typical uncertainties of normalised
magnitudes from plate digitisations, of no less than $\sim$0.3\,mag 
(Hambly et~al. 2001). 
These facts indicate that: ($i$) the large DENIS photometric uncertainties in
the optical (especially $\delta I$ = 1.00\,mag) are overestimated; and ($ii$)
the uncorrected DENIS magnitudes can still be used with uncertainties at the
0.26\,mag level, except for high-amplitude variables.
The latter objects have, however, extremely red $J-K_{\rm s}$ colours (of
2.14$\pm$0.03\,mag in the case of V603~Ori) and are easy to recognise among
fore- and background sources.
Therefore, the DENIS multiple detections barely affect my selection criterion.

%

\subsection{The Mayrit catalogue}
\label{thecatalogue}

Finally, Table~\ref{estrellas} tabulates the Mayrit designations, the
most accurate DENIS $I$-band and 2MASS $JHK_{\rm s}$ magnitudes and J2000
coordinates of the {338} cluster members and member candidates that
constitute the Mayrit catalogue. 
The last column indicates with a filled star (``$\star$'') if the object is a
confirmed cluster member with at least one known feature of~youth.
Of the {338} objects, only 16 have no cluster membership information
and are fainter than the completeness of the search; 
likewise, only 3 member candidates are fainter than the detection limit.
However, the numbers of confirmed cluster members fainter than the completeness
(43) and detection (15) limits are larger.
The correspondingly larger photometric uncertainties might explain the width of
the bottom of the cluster sequence. 

The catalogue has advantages and disadvantages.
On the one hand, its most noticeable pros are:
\begin{itemize} 
\item{comprehensiviness: tabulated Mayrit sources outnumbers
previous studies in $\sigma$~Orionis;}   
\item{very wide investigated magnitude interval: 
it translates into a very wide mass interval, that ranges four orders of
magnitude, from the 18+12\,$M_\odot$ of $\sigma$~Ori~AB to the
$\sim$0.03\,$M_\odot$ of {B05~2.03--671} (Caballero \& Chabrier,
in~prep.).  
The cluster stellar and substellar populations had been traditionally
studied in mass~bits;}  
\item{continuity and symmetry of the survey area: on the contrary to
previous searches in the cluster, there are no gaps between detectors or
asymmetric layouts.}  
\item{data homogeneity: only DENIS and 2MASS astrometric and photometric
data are tabulated for all the objects (except for a few bright
stars -- see Section~\ref{correlation});}
\end{itemize}  

On the other hand, some disadvantages of the Mayrit catalogue~are:
\begin{itemize} 
\item{extreme heterogenity of the list of confirmed cluster members: I~have
examined several dozens works, including spectroscopy, mid-infrared
photometry and X-ray emission, for making up the catalogue;} 
\item{moderate size (0.78\,deg$^2$): a search radius larger than 30\,arcmin
would lead to survey the Horsehead Nebula-IC~434 complex, where the extinction
is very high and background objects are reddened (just as in
Section~\ref{selection}). 
The variable extinction would make the cluster member selection to be very
difficult;} 
\item{incompleteness and contamination, that depend on the brightness
interval and the spatial distribution of young stars in the Ori~OB~1~b
association.
Both incompleteness and contamination of this catalogue will be discussed in a
future~work.} 
\end{itemize}

\section{Summary}

The $\sim$3\,Ma-old $\sigma$~Orionis cluster is a new cornerstone for
observational and theoretical studies with the aim to understand the general
processes of collapse and fragmentation of a molecular cloud, formation of stars
and substellar objects and evolution of circumstellar discs.
I~present a comprehensive, rather complete catalogue of cluster members
that can be used for further studies in $\sigma$~Orionis (e.g. spatial
distribution, multiplicity, mass function and frequency and characterisation
of~discs).  
This investigation covers the whole stellar and part of the brown dwarf domain
of the cluster from $\sim$18 to $\sim$0.03\,$M_\odot$.

I~have performed an $IK_{\rm s}$ survey in a 30\,arcmin-radius region
centred on the O9.5V+B0.5V binary $\sigma$~Ori~AB using the Aladin tool and
optical and near-infrared data from the DENIS and 2MASS catalogues.
The photometric data have been complemented with information from the literature
regarding the membership of known sources.
I~have compiled a list of 241 cluster stars and brown dwarfs with known
features of youth (e.g. early spectral types, Li~{\sc i} in absorption, near and
mid infrared excess attributed to surrounding discs, strong X-ray and H$\alpha$
emissions), 85 fore- and background stars with astrometric and spectroscopic
data (nine are new), {18} galaxies with extended FWHMs and 12 probable
reddened sources in a nebulosity northeastern of the survey area, associated to
the Horsehead Nebula. 
From the $I$ vs. $I-K_{\rm s}$ diagram, I have selected {97} additional
photometric cluster member candidates without reliable membership information.
This makes a list of {338} $\sigma$~Orionis members and member candidates, 
the Mayrit catalogue, of which more than 70\,\% display features of extreme
youth.  
I~tabulate precise coordinates, $IJHK_{\rm s}$ and suplementary
information of all the cluster members, member candidates and non-members.


\begin{acknowledgements}

I thank the anonymous referee, C.~A.~L.~Bailer-Jones, T.~Forveille
and R.~Mundt for helpful suggestions and discussion. 
Partial financial support was provided by the Universidad Complutense de
Madrid and the Spanish Ministerio Educaci\'on y Ciencia under grants 
AyA2004--00253 and AyA2005--02750 of the Programa Nacional de Astronom\'{\i}a y
Astrof\'{\i}sica and by the Comunidad Aut\'onoma de Madrid under PRICIT project
S--0505/ESP--0237 (AstroCAM). 

\end{acknowledgements}

\appendix

\section{Tables~\ref{spektra} to \ref{references}}


\begin{longtable}{llll}
\caption{\label{spektra} Confirmed young stars and brown dwarfs in $\sigma$ Orionis.}\\
\hline
\hline
            \noalign{\smallskip}
Name		& Alternative name(s)	& Feature(s) of youth 	& References \\  
            \noalign{\smallskip}
\hline
            \noalign{\smallskip}
\endfirsthead
\caption{Confirmed young stars and brown dwarfs in $\sigma$ Orionis (cont.).}\\
\hline
\hline
            \noalign{\smallskip}
Name		& Alternative name(s)	& Feature(s) of youth 	& References \\  
            \noalign{\smallskip}
\hline
            \noalign{\smallskip}
\endhead
\hline
\endfoot
Mayrit      AB  & \object{$\sigma$ Ori} A + B + IRS1, HD 37468  	& OB, mIR, XXX  		& FA04,MKK43,Fr06,Ca07a \\ 
Mayrit   42062  & \object{$\sigma$ Ori E}, V1030 Ori			& OB, XXX			& GW58,Le68,HHG89,Ca07a \\   
Mayrit   13084  & \object{$\sigma$ Ori D}, HD 37468D			& OB, X?			& CP18,Ca07a,Ca07b \\ 
Mayrit 1548068  & \object{HD 37699}, 1AXG J054020--0225 		& OB, mIR, XX			& Ne43,Gr92,Ca07a \\		   
Mayrit  524060  & \object{HD 37564}, BD--02 1330			& OB, debris disc, X		& TO91,Ca07a,He07 \\		   
Mayrit  208324  & \object{HD 294271}, BD--02 1324			& OB				& SC71,Gu81,Ca07a \\	 
Mayrit  306125  & \object{HD 37525} AB, BD--02 1328			& OB, XX			& SC71,Ni76,Fr06,Ca07a \\	
Mayrit  182305  & \object{HD 294272 A}, BD--02 1323			& OB, X				& SC71,Fr06,Ca07a \\
\end{longtable}


\begin{longtable}{lllcll}
\caption{\label{NM} Fore- and background stars.}\\
\hline
\hline
            \noalign{\smallskip}
2MASS			& Alternative  			& Sp.	& $\mu_\alpha \cos{\delta},\mu_\delta$ 	& Remarks	& References 	\\  
designation		& name(s)			&	& (mas\,a$^{-1}$)			&		&  		\\  
            \noalign{\smallskip}
\hline
            \noalign{\smallskip}
\endfirsthead
\caption{Fore- and background stars (cont.).}\\
\hline
\hline
            \noalign{\smallskip}
2MASS			& Alternative  			& Sp.	& $\mu_\alpha \cos{\delta},\mu_\delta$ 	& Remarks	& References 	\\  
designation		& name(s)			&	& (mas\,a$^{-1}$)			&		&  		\\  
           \noalign{\smallskip}
\hline
            \noalign{\smallskip}
\endhead
\hline
\endfoot
J05382849--0303338	& \object{HD 294280}		& Yes   & $\lesssim$ 10 	& foreground K5		 		& Ne95,Ca07a		\\ 
J05375734--0253177	& \object{HD 294277}		& Yes	& $\lesssim$ 10		& foreground K2 		      	& Ne95,Ca07a	       	\\ 
J05383876--0249013	& \object{HD 294278}		& Yes	& +10.9,+2.6		& foreground K2, no $\mu$	       	& Ne95,Ca07a	       	\\ 
J05372988--0243458	& \object{TYC 4771 1468 1}	& No	& $\lesssim$ 10		& foreground K--M:		       	& Ca07a  	       	\\		  
J05365714--0225400	& \object{TYC 4770 1018 1}	& No	& $\lesssim$ 10		& foreground K--M:		       	& Ca07a  	       	\\		  
J05391394--0210493	& \object{TYC 4771 1012 1}	& No	& $\lesssim$ 10		& foreground K--M:		       	& Ca07a  	       	\\		  
J05394306--0228456	& \object{TYC 4771 934 1}	& No	& $\lesssim$ 10		& foreground K--M:		       	& Ca07a  	       	\\		  
J05400217--0253423  	& ...				& No	& $\lesssim$ 10		& foreground M:				& ...			\\
\end{longtable}

   \begin{table*}
      \caption[]{Galaxies in the survey area.}  
         \label{galaxies}
     $$ 
         \begin{tabular}{llcccll}
            \hline
            \hline
            \noalign{\smallskip}
2MASS			& Alternative  	        	& $I \pm \delta I$      & $J \pm \delta J$	& $K_{\rm s} \pm \delta K_{\rm s}$ & Remarks 	& References 	\\  
designation		& name(s)	        	& (mag) 	        & (mag) 		& (mag) 		&			& 		\\  
            \noalign{\smallskip}
            \hline
            \noalign{\smallskip}
J05380008--0223516	& ...				& 14.711 $\pm$ 1.00	& 13.441 $\pm$ 0.077	& 12.301 $\pm$ 0.072	& ... 		    	& 2MX03	    	\\
J05370717--0241453   	& ...				& 15.729 $\pm$ 1.00	& 14.465 $\pm$ 0.087	& 13.172 $\pm$ 0.078	& ...			& 2MX03 	\\
J05380720--0303036   	& ...				& 16.102 $\pm$ 1.00	& 14.559 $\pm$ 0.076	& 13.319 $\pm$ 0.073	& ...			& 2MX03 	\\
J05372792--0240033 	& \object{2E 1448}	 	& 16.186 $\pm$ 1.00	& 14.825 $\pm$ 0.078	& 13.432 $\pm$ 0.074	& XX 			& 2MX03,He07,CL-S \\ 
J05373789--0240544   	& ...				& 15.982 $\pm$ 1.00	& 14.840 $\pm$ 0.071	& 13.560 $\pm$ 0.070	& ...			& 2MX03 	\\
J05382931--0233125	& ...				& 16.275 $\pm$ 0.07	& 14.853 $\pm$ 0.087	& 13.713 $\pm$ 0.089	& ...			& 2MX03 	\\
\hline
         \end{tabular}
     $$ 
   \end{table*}

   \begin{table*}
      \caption[]{Probable reddened sources in the northeastern
      nebulosity.} 
         \label{northernrag}
     $$ 
         \begin{tabular}{llcccll}
            \hline
            \hline
            \noalign{\smallskip}
2MASS			& Alternative  	        & $I \pm \delta I$      & $J \pm \delta J$	& $K_{\rm s} \pm \delta K_{\rm s}$ & Remarks 	& References 	\\  
designation		& name(s)	        & (mag) 	        & (mag) 		& (mag) 		&			& 		\\  
            \noalign{\smallskip}
            \hline
            \noalign{\smallskip}
J05392159--0212171	& ...		        & 14.191 $\pm$ 0.16     & 12.092 $\pm$ 0.026	& 11.360 $\pm$ 0.019	& blue $J-K_{\rm s}$	& ...	       \\ 
J05393293--0211312  	& \object{SWW 93}       & 15.636 $\pm$ 0.24     & 13.011 $\pm$ 0.024	& 12.055 $\pm$ 0.021	& Haro 5--24?		& Sh04         \\ 
J05390830--0207002   	& ...  		        & 15.389 $\pm$ 0.24     & 13.234 $\pm$ 0.026	& 12.543 $\pm$ 0.026	& blue $J-K_{\rm s}$	& ...	       \\ 
J05390597--0207313  	& ...		        & 16.536 $\pm$ 0.24     & 13.656 $\pm$ 0.028	& 12.572 $\pm$ 0.023	& ...			& ...	       \\ 
J05385224--0208098 	& \object{SWW 110}      & 15.644 $\pm$ 0.05     & 13.692 $\pm$ 0.026	& 12.818 $\pm$ 0.026	& ...			& Sh04         \\ 
J05391653--0207331 	& \object{SWW 96}       & 16.035 $\pm$ 0.24     & 13.826 $\pm$ 0.027	& 12.967 $\pm$ 0.034	& ...			& Sh04         \\
\hline
         \end{tabular}
     $$ 
   \end{table*}
%

\begin{longtable}{llll}
\caption{\label{new} Candidate young stars and brown dwarfs in $\sigma$
Orionis.}\\ 
\hline
\hline
            \noalign{\smallskip}
Name		& Alternative name(s)	& Remarks 	& References \\  
            \noalign{\smallskip}
\hline
            \noalign{\smallskip}
\endfirsthead
\caption{Candidate young stars and brown dwarfs in $\sigma$ Orionis (cont.).}\\
\hline
\hline
            \noalign{\smallskip}
Name		& Alternative name(s)	& Remarks 	& References \\  
            \noalign{\smallskip}
\hline
            \noalign{\smallskip}
\endhead
\hline
\endfoot
Mayrit 1165138  & ...			& New, G:-type		& ...		\\ 
Mayrit  240322  & \object{BD--02 1324D} & ...			& ...		\\ 
Mayrit  882239  & ...			& New, G:-type		& ...		\\ 
Mayrit 1338116  & ...			& New			& ...		\\ 
Mayrit  123000  & \object{P053844--0233} & ...			& Wo96,Ca07b	\\ 
Mayrit 1045356  & ...			& New			& ...		\\ 
Mayrit 1564345  & ...			& New			& ...		\\ 
Mayrit 1169117  & ...			& New			& ...		\\
\end{longtable}

\begin{longtable}{lccccccc}
\caption{\label{estrellas} The Mayrit catalogue: J2000 coordinates and $IJHK_{\rm s}$-band
      magnitudes of $\sigma$ Orionis members and candidate members.}\\
\hline
\hline
            \noalign{\smallskip}
Name			& $\alpha$  	& $\delta$ 	& $I \pm \delta I$	& $J \pm \delta J$	& $H \pm \delta H$	& $K_{\rm s} \pm \delta K_{\rm s}$ 	& Young \\ 
			& (J2000)  	& (J2000)  	& (mag)  		& (mag)  		& (mag)  		& (mag)  				& \\  
            \noalign{\smallskip}
\hline
            \noalign{\smallskip}
\endfirsthead
\caption{The Mayrit catalogue: J2000 coordinates and $IJHK_{\rm s}$-band
      magnitudes of $\sigma$ Orionis members and candidate members (cont.).}\\
\hline
\hline
            \noalign{\smallskip}
Name			& $\alpha$  	& $\delta$ 	& $I \pm \delta I$	& $J \pm \delta J$	& $H \pm \delta H$	& $K_{\rm s} \pm \delta K_{\rm s}$ 	& Young \\ 
			& (J2000)  	& (J2000)  	& (mag)  		& (mag)  		& (mag)  		& (mag)  				& \\  
            \noalign{\smallskip}
\hline
            \noalign{\smallskip}
\endhead
\hline
\endfoot
            \noalign{\smallskip}
\object{Mayrit      AB}& 05 38 44.76   & --02 36 00.2  &  4.02  $\pm$ 0.01     &  4.330 $\pm$ 0.010    &  4.500 $\pm$ 0.010    &  4.490 $\pm$ 0.016 & $\star$	\\ 
\object{Mayrit   42062}& 05 38 47.21   & --02 35 40.5  &  6.86  $\pm$ 0.10     &  6.974 $\pm$ 0.026    &  6.954 $\pm$ 0.031    &  6.952 $\pm$ 0.029 & $\star$	\\ 
\object{Mayrit   13084}& 05 38 45.62   & --02 35 58.9  &  6.98  $\pm$ 0.03     &  7.116 $\pm$ 0.029    &  7.219 $\pm$ 0.027    &  7.260 $\pm$ 0.021 & $\star$	\\ %
\object{Mayrit 1548068}& 05 40 20.19   & --02 26 08.2  &  7.77  $\pm$ 0.10     &  7.841 $\pm$ 0.029    &  7.904 $\pm$ 0.042    &  7.880 $\pm$ 0.031 & $\star$	\\ %
\object{Mayrit  524060}& 05 39 15.06   & --02 31 37.6  &  8.18  $\pm$ 0.10     &  7.976 $\pm$ 0.021    &  7.881 $\pm$ 0.038    &  7.828 $\pm$ 0.029 & $\star$	\\ 
\object{Mayrit  208324}& 05 38 36.55   & --02 33 12.8  &  7.99  $\pm$ 0.10     &  8.100 $\pm$ 0.021    &  8.180 $\pm$ 0.055    &  8.202 $\pm$ 0.038 & $\star$	\\ %
\object{Mayrit  306125}& 05 39 01.49   & --02 38 56.4  &  8.21  $\pm$ 0.10     &  8.131 $\pm$ 0.030    &  8.105 $\pm$ 0.042    &  8.093 $\pm$ 0.020 & $\star$	\\ 
\object{Mayrit  182305}& 05 38 34.80   & --02 34 15.8  &  8.40  $\pm$ 0.10     &  8.346 $\pm$ 0.026    &  8.380 $\pm$ 0.046    &  8.374 $\pm$ 0.026 & $\star$	\\
\end{longtable}

\begin{longtable}{lr}
\caption{\label{references} Alphanumeric abbreviations for
references used in Tables~\ref{spektra} to \ref{new}.}\\
\hline
\hline
            \noalign{\smallskip}
Abbreviation	& Reference 			\\ %
            \noalign{\smallskip}
\hline
            \noalign{\smallskip}
\endfirsthead
\caption{Alphanumeric abbreviations for
references used in Tables~\ref{spektra} to \ref{new} (cont.).}\\
\hline
\hline
            \noalign{\smallskip}
Abbreviation	& Reference 			\\ %
            \noalign{\smallskip}
\hline
            \noalign{\smallskip}
\endhead
\hline
\endfoot
FA04		& Frost \& Adams (1904) 	\\ %
CP18		& Cannon \& Pickering (1918--1924) \\ %
MKK43		& Morgan, Keenan \& Kellman (1943) \\ %
Ne43		& Neubauer (1943) 		\\ %
HM53		& Haro \& Moreno (1953) 	\\ %
GW58		& Greenstein \& Wallerstein (1958) \\ %
GBT61		& Giclas, Burnham \& Thomas (1961) \\ %
Le68		& Lesh (1968) 			\\
\end{longtable}

\end{document}